\documentclass[12pt]{article}
\usepackage{a4wide,epsfig,cite}

\newlength {\pbreite}
\newlength {\parbreite}
\newlength {\kbreite}
\newlength {\lbreite}
\newlength {\qbreite}
\newcommand{\btau}{\mbox{\boldmath{$\tau$}\unboldmath}}

\newcommand{\bpi}{\mbox{\boldmath{$\pi$}\unboldmath}}
\newcommand{\bphi}{\mbox{\boldmath{$\phi$}\unboldmath}}

\newcommand{\bx}{{\bf x}}
\newcommand{\bX}{{\bf X}}

\newcommand{\bn}{{\bf n}}
\newcommand{\bM}{{\bf M}}
\newcommand{\by}{{\bf y}}
\newcommand{\bz}{{\bf z}}

\newcommand{\bk}{{\bf k}}
\newcommand{\be}{\begin{equation}}
\newcommand{\ee}{\end{equation}}
\newcommand{\bea}{\begin{eqnarray}}
\newcommand{\eea}{\end{eqnarray}}
\newcommand{\pkt}{\; .}
\newcommand{\kma}{\; ,}
\newcommand{\sub}[1]{_{\rm #1}}
\newcommand{\eqn}[1]{(\ref{#1})}
\newcommand{\calv}{{\cal V}}
\newcommand{\bgamma}{\mbox{\boldmath{$\gamma$}\unboldmath}}
\newcommand{\balpha}{\mbox{\boldmath{$\alpha$}\unboldmath}}  

\newcommand{\bnabla}{\mbox{\boldmath{$\nabla$}\unboldmath}}

\newcommand{\hbk}{\mbox{\boldmath{$\hat k$}\unboldmath}}
\newcommand{\hbz}{\mbox{\boldmath{$\hat z$}\unboldmath}}

\newcommand{\hbx}{\mbox{\boldmath{$\hat x$}\unboldmath}}

\newcommand{\bsigma}{\mbox{\boldmath{$\sigma$}\unboldmath}}
\newcommand{\ebx}{\mbox{\boldmath{$\scriptstyle x$}\unboldmath}}

\newcommand{\ebk}{\mbox{\boldmath{$\scriptstyle k$}\unboldmath}}

\newcommand{\he}{h_l^{(1)}}

\renewcommand{\Im}{{\rm Im\,}}
\newcommand{\tr}{{\rm tr\,}}
\newcommand{\Tr}{{\rm Tr\,}}
\begin{document}
\begin{titlepage}
\begin{flushright}
DO-TH-00/17\\
November 2000
\end{flushright}

\vspace{20mm}
\begin{center}
{\Large \bf Parton distributions in the chiral quark model: \\
a continuum computation }

\vspace{10mm}

{\large  J\"urgen Baacke\footnote{
e-mail:~baacke@physik.uni-dortmund.de} and 
Hendrik Sprenger\footnote{e-mail:~sprenger@hal1.physik.uni-dortmund.de} } \\
{\large Institut f\"ur Physik, Universit\"at Dortmund} \\
{\large D - 44221 Dortmund , Germany}
\\
\vspace{8mm}
\bf{Abstract}
\end{center}
We compute the parton distributions for the chiral quark model.
We present a new technique for performing such computations
based on Green functions. This approach avoids a discretization
of the spectrum. It therefore does not need any smoothing procedures.
 The results are similar to those of other
groups, however the distributions peak at smaller $x$. 
\vspace{1cm}

\noindent
PACS: 12.38.Lg, 12.39.Ki, 11.15.Pg, 14.20.Dh, 13.60.Hb

\end{titlepage}


\section{Introduction}
\setcounter{equation}{0}

Nucleon structure functions  have been
measured since three decades, yielding very precise and detailed
information about the parton distributions. Their evolution with 
$Q^2$ is one of the important tests of QCD \cite{Reya:1981zk}
and the predicted asymptotic freedom.
There are various phenomenogical ans\"atze for the form of the
parton distribution, which necessarily has to be chosen at some
fixed value of $Q^2$. In principle these functions contain
important information on the structure of the nucleon, which
in turn should yield information on the dynamics that determines 
this structure. The problem is that at low $Q^2$ where one should 
expect the nucleon to be described by quark model wave functions
QCD becomes nonperturbative so that a description by gluons and 
current quarks is no longer accessible. Rather one expects the
system to be described by other effective field degrees of freedom.
Among various models which are used to describe the 
low energy properties like form factors,  couplings to 
the pion and electromagnetic fields or resonance excitations
the chiral quark model \cite{Alkofer:1996} contains several elements a realistic model
should have: there are  constituent quarks, there are mesons,
the model has chiral symmetry. The nucleon is described as a chiral
hedgehog with an occupied valence quarks bound state, 
as an extension of the Skyrme model.
The fact that the model contains quarks, and that the meson field
is not elementary, but itself a quark-antiquark condensate,
implies that the coupling of the electromagnetic current to
the nucleon is described entirely by the quark degrees of freedom.
This property makes it a promising issue to compute the nucleon
structure functions in this model. 

This has been done  recently  
by several groups \cite{Diakonov:1996sr,Diakonov:1997vc,
Wakamatsu:1998en,Weigel:1996ef}. We do not have anything to add to the general
approach and its basic formulae relating the parton distribution
to the quark mode functions or the quark Green function in the 
external field of the chiral hedgehog. 
We also make a number of  technical  assumptions which have 
emerged to be necessary. So the pion field is varied
only on the chiral circle, the problem of translation
variance remains,  and the Pauli-Villars subtraction
with a finite cutoff is applied to the quark sea, not to the
valence quarks. This is done consistently as well  for the 
self-consistent nucleon solutions as also for the
parton distributions \cite{Weiss:1997rt}.
 
Where we depart from the
previous approaches is in the method used for the
practical computation of the sea quark distribution functions. 
All previous approaches start by discretizing the mode spectrum,
introducing a large ``confining'' sphere with appropriate
boundary conditions for the quark wave functions.
As the spectrum is discrete, the parton distributions are not smooth,
and they have to be modified by a smoothing procedure.
Though this, and the discretization as such, could introduce some 
arbitrariness the various groups obtain similar, though
slightly different, results.

It  has been  demonstrated previously that such a
discretization can be
avoided entirely  when computing the energy of the ``quark sea''. 
This quantity has been 
computed by Moussallam \cite{Moussallam:1989uk} using phase shifts, 
an approach well-known
from kink quantization \cite{Rajaraman:1989}.
Alternatively \cite{Baacke:1990sb} the self energy
can be computed from the Euclidian Green function.
A self-consistent computation of the hedgehog profile for 
the nucleon was carried óout recently \cite{Baacke:1998nm}. 
We will present here a technique for computing the parton distribution
in a continuum approach, i.e., without any discretization.

While the self energy can be computed using the Euclidean
Green function, by introducing and deforming a suitable
integration contour in the complex energy plane, we have to use
the Green function in Minkowski space-time if we want to
evaluate the parton distribution. So instead of working  far from
the physical cuts of the Green function, we have to work on the 
cut, here. The technique we use  is  similar
to Moussallam' s approach. But of course we compute different
physical quantities. So we need more information than
just the scattering phase shifts. 
Furthermore  we will make use of the techniques
well-known in potential scattering \cite{deAlfaro:1965} 
in order to find 
expressions for the $S$ matrix elements that are suitable both for
discussing their analytic properties and regularization, and for numerical 
computation. Besides avoiding discretization another advantageous feature
of our approach is the inclusion of the asymptotic spectrum.
Indeed the power behavior of the asymptotic tails of the angular 
momentum summation and of the momentum integrations is known and
these tails can be included on the basis of fits.

The plan of the paper is as follows:
In section 2 we introduce the model and the 
expression which relates the parton distribution to the
quark Green function. A short derivation of this relation
is given in Appendix A. 
In section 3 we find analytic expressions for the Green function
and for the $S$ matrix, for the case of a bosonic quantum field
in an external field. The corresponding expressions for
the fermions, used in the real calculation, are given in
Appendix B. In Appendix C we derive a relation for the Fourier 
transform of the Green function, which is the basis for
the numerical calculation.
The numerical procedure is described in section 4, in
section 5 we discuss the results and present our conclusions.


\section{The model}
\setcounter{equation}{0}
The Nambu-Jona-Lasinio (NJL) model is defined by the
Lagrangian \cite{Nambu:1961a}
\begin{equation}
{\cal L}_{\rm NJL}=
{\bar{\psi}}(i\gamma^{\mu}\partial_{\mu}-m)\psi+ \frac{G}{2}
\left[(\bar{\psi}\psi)^2+(\bar{\psi}i\gamma_5{\btau}\psi)^2\right]
\pkt
\end{equation}
Introducing Lagrange multiplier fields for the scalar and pseudoscalar 
densities, the action can be written in bosonized form as
\begin{equation} \label{boson}
S_{\rm NJL}=\int  d ^4 x\Biggl\{ \bar{\psi}
\biggl[i\gamma^{\mu}\partial_{\mu}-g\Bigl(\sigma+i{\bpi}\cdot{\btau}
\gamma_5\Bigr)\biggr]
\psi-\frac{\mu^2}{2}\left(\sigma^2+{\bpi}^2\right)+
\frac{m\mu^2}{g}\sigma\Biggr\}\pkt
\end{equation}
The parameters of the model are the fermion self-coupling $G$, the quark 
mass $m$. Furthermore, as the model is non-renormalizable, a cutoff
$\Lambda$ has to be introduced. In the bosonized 
version these parameters appear as the quark-meson coupling $g$ and the
symmetry breaking mass parameter $\mu$. They are related to the basic
parameters as
$g=\mu \sqrt{G}$ and $ m\mu^2=gf_\pi m_\pi^2$.
The latter equation expresses $\mu$ in terms of 
physical constants and of the coupling $g$ which remains a
free parameter. A further
relation is obtained from the gradient expansion of
the effective action. The resulting kinetic 
term of the pion field is normalized correctly if the 
Pauli-Villars cutoff is fixed as
\begin{equation}
\Lambda=M\sqrt{\exp\left(\frac{4\pi^2}{N_{c} g^2}\right)}\pkt
\label{lambda}
\end{equation}
Here $M$ is the ``dynamical quark mass'' $M=g f_\pi$.

We will not consider the explicit breaking of chiral symmetry
by the ``current'' quark mass $m$, so we work in the limit
$m_\pi=0$. Furthermore we will restrict 
the variation 
of $\sigma$ and $\pi$ to the chiral hypersphere
\be
\sigma^2+ \bpi^2 = f_\pi^2
\,.\ee
The the second
term in Eq.~\ref{boson} becomes a constant that can be
chosen to be zero.
In this case, as  shown by \cite{Reinhardt:1988fz}
 in proper time regularization,
and in \cite{Doring:1992sj}
 in the Pauli-Villars regularization used here, 
the model admits a solution which can be identified as the nucleon.

The self-consistent nucleon configuration  is based on  a chiral hedgehog
ansatz 
\be
g\left[\sigma+i\gamma_5{\btau}\cdot{\bpi}\right]
=M\exp\left\{i\gamma_5{\btau}\cdot{\bphi}(\bx)
\right\}\equiv \bM(\bx)
\ee
with 
\be
{\bphi}(\bx)=\hbx\vartheta(r)
\ee
for the $\sigma$ and $\pi$ fields. In this external
field the quark field has a bound state solution identified
with the valence quarks. Furthermore the external field
leads to a modification of the ``sea quark'' energy.
The self-consistent solution is
obtained by minimizing the energy 
\be\label{SF}
E_{\rm quark}=\frac{1}{\tau}\Tr\log\Bigl[{-i\gamma^{\mu}\partial_{\mu}
+\bM(\bx)}\Bigr]\,,
\ee
where the trace extends over the negative energy levels as well as
over the bound state. It is understood that the vacuum energy is 
subtracted. Furthermore a Pauli-Villars subtraction using the
regulator $\Lambda$ is required.  

We use here  profiles $\vartheta(r)$ obtained by us previously 
\cite{Baacke:1998nm}. There the sea quark energy was
obtained as a trace over Euclidean Green functions.

It has been noticed by several authors that the computation
of the parton distributions can be related to the Green
function of the quarks in the external field. In contrast to
the computation of the energy we have to use the Green
function at real energy, not the Euclidean one. 
This requires some changes with respect to  our previous approach, 
but we will retain its benefits: to avoid working with 
discrete levels and the corresponding wave functions.  

We rederive the expression for the parton distribution
in Appendix A, following \cite{Diakonov:1996sr}, we obtain for the sea quark
contribution
\begin{eqnarray}
\label{DiaS}
D_i(x)&=&-\Im N_{c}M\sub{N}\int\frac{d^4 k}{(2\pi)^4}\delta(M\sub{N}x-k_3-k_0)\nonumber\\
&&\times\Theta(-k_0-M)
\tr\Bigl[\gamma_0(1+\gamma_0\gamma_3)
S_{\rm F}(\bk,\bk,k_0)\Bigr]\pkt
\end{eqnarray}
For the antiquarks one has to replace $x \to -x$ and
to reverse the sign of the argument on the $\Theta$ function.
Here and in the following $M\sub{N}$ denotes the nucleon mass, $x$
is the Bjorken scaling variable, and $S_{\rm F}(\bk,\bk,k_0)$
is the Green function of the fermions in the external
hedgehog field.
It is the Fourier transform of the position space Green
function $S_{\rm F}(\bx,\by,t)$ which satisfies
\be
\left[i\gamma_{\mu}\partial^{\mu}_x
-\bM(\bx)\right] S_{\rm F}(\bx,\by,t-t')=\delta(\bx-\by)\delta(t-t')
\ee
with Feynman boundary conditions. As the external 
hedgehog field is not translationally invariant
it depends on two separate spatial variables, and so does its Fourier
transform
\be \label{greenfou}
S_{\rm F}(\bk,\bk',k_0)=\int dt d^3 x d^3 y
e^{i\left(k_0 t -\bk\cdot\bx + \bk'\cdot \by \right)}
S_{\rm F}(\bx,\by,t)\pkt
\ee
The relation to the positive and negative energy 
eigenfunctions $U_\alpha$, resp., $V_\alpha$ is formally given by
\bea \nonumber
S_{\rm F}(\bx,\by,t)&=&-i\langle T(\psi(\bx,t)\bar\psi(\by,0)\rangle
 \\ \label{greenuv}
&=&i\Theta(t)\sum_{\alpha > 0}e^{-iE_\alpha t}U_\alpha(\bx)\bar
U_\alpha(\by)
\\\nonumber
&& -i\Theta(-t)\sum_{\alpha < 0}e^{i|E_\alpha| t}V_\alpha(\bx)\bar
V_\alpha(\by)\pkt
\eea
Here $\alpha > 0$ ($\alpha < 0$) symbolizes the positive (negative)
energy states. We will not use this definition for the
practical computation, as it would necessitate the discretization
of the energy spectrum.
Rather the Green function is obtained in terms of mode functions
as described in \cite{Baacke:1998nm}.

As usual the Green function $ S\sub{F}$ is written in terms
of a bosonic Green function using the ansatz
 \be
   S\sub{F}=(E+H)G\,
 \ee
where $G$ satisfies
\be\label{DGL2}
\left(H^2-E^2\right)G(\bx,\bx',E)=
\left[-\Delta+M^2+{\cal{V}}(\bx)-E^2\right]
G(\bx,\bx',E)=-\delta^3(\bx-\bx')
\pkt\ee
Here $\calv$ is the  potential or vertex operator 
\be \label{potential}
  {\cal{V}(\bx)}=i\bgamma\cdot\bnabla\bM(\bx)\,.
\ee
 $G(\bx,\bx',E)$ is the inverse
of a symmetric operator, it is obtained by standard techniques
described in Appendix B.

In addition to the sea quarks the valence quarks  contribute
to the structure functions. The equation for the  $K^P=0^+$
partial wave is 
\begin{equation}
\label{EWgl}
\Bigl[-\Delta+{\cal V}^{0^+}(\bx)\Bigr]\psi_0=\omega_0^2\psi_0 \pkt
\end{equation}
 The spinor $\psi_0$ is determined by
two radial wave functions $h(r)$ and $j(r)$, and the potential
${\cal V}^{0+}$ is a $2\times 2$ matrix given in \cite{Baacke:1998nm}.
Defining their Fourier transforms as
\begin{eqnarray}
h(k)&=&\int_0^\infty d r r^2 j_0(kr)h(r)\kma\\
j(k)&=&\int_0^\infty d r r^2 j_1(kr)j(r)
\end{eqnarray}
one finds 
\begin{equation}
\label{Diunpol}
D_{i}^{\rm bou}(x)=\frac{N\sub{c} M\sub{N}}{\pi}\int_{|M\sub{N}x-E_{\rm bou}|}^\infty d  k \,k
\left(h(k)^2+j(k)^2-2\frac{M\sub{N}x-E_{\rm bou}}{k}h(k)j(k)\right)\pkt
\end{equation}
This expression is equivalent to the bound-state in
\cite{Diakonov:1996sr}. The bound state distribution function is convergent. 


\section{The Green function }
\setcounter{equation}{0}
We present  here an explicit
expression for the Green function that can be evaluated without
having recurse to discretization, we also derive some
further  identities that are useful in this context.
For the sake of transparency
we here consider a scalar field in a spherically symmetric
background. The much more involved expressions for fermion
fields coupled to the chiral hedgehog are presented in Appendix B.

We consider a scalar field with a space 
dependent  mass term $m(\bx)$ which for $|\bx|\to \infty$
tends to $M$. We furthermore assume that $m(\bx)$ only depends
on $|\bx|=r$
We decompose 
\be
m^2(\bx)= M^2 +\calv(r)\pkt
\ee
The Green function then obeys the differential equation
\begin{equation} 
\label{Greendgl}
\left[-E^2 - \Delta +\calv(r) +M^2\right]
G(\bx,\bx';E)=-\delta^3(\bx-\bx')\pkt
\end{equation}
The  Fourier transformation of the {\em free} \/Green 
function is given by
\begin{equation}
G^{(0)}(\bx,\bx';E)=
\int\frac{d^3k}{(2\pi)^3}\frac{e^{i\ebk(\ebx-\ebx')}}
{\kappa ^2-k^2+i\epsilon}\,,
\end{equation}
where  $\kappa ^2=E^2-M^2$. 
As is well-know the $k$ integration can be carried
out and one obtains 
\be
G^{(0)}(\bx,\bx';E)=
-\frac{\kappa }{4\pi}\frac{e^{i(\kappa +i\epsilon)R}}{\kappa R}
\kma
\ee
with $R=\sqrt{r^2+r^{'2}- 2 r r' \cos(\theta)}$.
It can be decomposed with respect to Legendre polynomials  using the Gegenbauer
expansion.
One finds
\bea
\label{freex}
G^{(0)}(\bx,\bx';E)&=&
\kappa 
\sum_l \frac{(2l+1)}{4\pi}j_l(\kappa r_{_<})
\Bigl[y_l(\kappa r_{_>})-ij_l(\kappa r_{_>})\Bigr]P_l(\cos(\theta))
\nonumber\\  &=&
-i\kappa 
\sum_l \frac{(2l+1)}{4\pi}j_l(\kappa r_{_<})
h_l^{(1)}(\kappa r_{_>})P_l(\cos(\theta))\pkt
\eea
The spherical Bessel functions are defined as in \cite{Abramowitz:1972}.
The imaginary part is given by
\begin{equation}
\label{Quadrat}
\Im G^{(0)}(\bx,\bx';E)=
-\kappa  \sum_l \frac{(2l+1)}{4\pi}j_l(\kappa r)j_l(\kappa r')P_l(\cos(\theta))\pkt
\end{equation}
The Green function in the external field is decomposed in an analogous way
via
\begin{equation}
 \label{exacfin}
G(\bx,\bx';E)=
-i\kappa 
\sum_l \frac{(2l+1)}{4\pi}
f_l^-(\kappa,r_{_<})f_l^+(\kappa,r_{_>})P_l(\cos (\theta))\pkt
\end{equation}
Here the radial mode functions $f_l^\pm$ satisfy the 
differential equation
\begin{equation}
\label{PartDGL}
\left(-\frac{d^2}{d r^2}-\frac{2}{r}\frac{d}{d r}
+\frac{l(l+1)}{r^2}-\kappa ^2\right)f_l(\kappa, r)=-{\cal{V}}(r)f_l(\kappa,r)\pkt
\end{equation}
The boundary conditions are, in analogy to the spherical Bessel functions:
\bea \label{scalarbc} 
\lim_{r\to 0} f^-(\kappa,r)/j_l(\kappa r)&=&\, {\rm const.}\kma
\\ \nonumber
\lim_{r\to \infty}f^+(\kappa,r)/h_l^{(1)}(\kappa r)&=& 1 \pkt
\eea
The normalization of $f_l^-(\kappa,r)$ is determined by the
Wronskian
\be 
r^2 W(f^+,f^-)=r^2\left(f_l^{+}\frac{d}{d r}f_l^{-}-
f_l^{-}\frac{d}{d r}f_l^{+}\right)=
\frac{-i}{\kappa}
\pkt
\ee
It is useful for the numerical computation as well as for the 
separation (subtraction) of the free Green function to
split off the Bessel functions via
\bea
\label{general1}
f^+_l(\kappa,r)&=&\left[1+c^+_l(\kappa,r)\right]h_l^{(1)}(\kappa r)\kma
\\
\label{general2}
f^-_l(\kappa, r)&=&\left[1+c_l^-(\kappa,r)\right]j_l(\kappa r)\pkt
\eea
The second equation is used only below the first zero
of $j_l(\kappa r)$. In fact the function $c_l^-(\kappa,r)$
is pathological as it has poles and zeros at the zeros
of $j_l$ and $f^-_l$, respectively.
An alternative, more useful expression for $f_l^-$ is obtained
by forming a linear combination of $f_l^+$ and $(f_l^+)^*$
which becomes regular as $r \to 0$; explicitly
\begin{eqnarray}
\label{first}
f_{l}^{-}(\kappa ,r)&=&\frac{1}{2}
\Biggl\{\nonumber\left(1+\bar{c}_{l}^{+}(\kappa ,r)\right)
h^{(2)}_{l}(\kappa r)
\\
&&+
\frac{1+{\bar c}_l^{+}(\kappa,0)}
{1+{c}_l^{+}(\kappa,0)}
\left(1+{c}_{l}^{+}(\kappa ,r)\right)
h^{(1)}_{l}(\kappa r)
\Biggr\}\pkt
\end{eqnarray}
The function $c_l^+(\kappa,r)$ tends to zero as $r \to \infty$
and goes to a constant as $r\to 0$. It is well suited for
numerical computation, using the differential equation
obtained by inserting the ansatz~\eqn{general1} into
Eq.~\eqn{PartDGL}.

The function $c^+_l$ is related in a simple way to
the $S$ matrix. The relation follows from considering the
regular solution $f_l^-$ in the form~\eqn{first}.
Indeed the $S$ matrix 
is defined via the asymptotic
behavior of the regular solution:
\begin{eqnarray}
\label{end2}
 f_l^-(\kappa ,r)
&\stackrel{r\to\infty}{\simeq}&\frac{1}{2}\left[
h_l^{(2)}(\kappa r)+
e^{2 i \delta_l(E)}h_l^{(1)}(\kappa r)
\right]
\\
&=&\frac{1}{2}\left[
h_l^{(2)}(\kappa r)+
\frac{1+\bar c_l^+(\kappa ,0)}
{1+ c_l^+(\kappa ,0)}
h_l^{(1)}(\kappa r)
\right]\nonumber \pkt
\end{eqnarray}
So the $S$ matrix is given by
\be
S_l(E)=e^{2 i \delta_l(E)}=\frac{1+\bar c_l^+(\kappa ,0) }{1+ c_l^+(\kappa ,0) }
\pkt
\ee
As is well-known \cite{Rajaraman:1989}, the phase shift can be used to compute
the zero point energy. For the quark sea in the chiral quark model
this has been done in \cite{Moussallam:1989uk};
the relation is, in this case,
\begin{equation}
E_0=-\frac 1 \pi \sum_{K^P}(2 K+1)\int_0^\infty d \kappa 
{\sqrt{\kappa^2+M^2}}
\frac{d \delta_K(\kappa)}{d \kappa} \pkt
\end{equation}

For the parton distribution we need the imaginary part of the Green
function in momentum space at equal three-momentum.
We have
\be
\Im G(\bk,\bk,E)=
\int d^3x\int d^3x'e^{-i\bk(\bx-\bx')}\frac{1}{2}
\left[G(\bx,\bx',E)-\bar G(\bx',\bx,E)\right]
\pkt
\ee
As the Green function is a symmetric in its spatial arguments
the expression in brackets is just twice the imaginary
part of the Green function in position space.
It is straightforward to show that this imaginary part is
given by
\begin{equation}
\Im G(\bx,\bx',E)=
-\kappa 
\sum_l \frac{(2l+1)}{4\pi} f_l^-(\kappa,r) 
\bar f^-_l(\kappa,r)P_l(\cos \theta)\pkt
\end{equation}
This is analogous to the imaginary part of the free Green 
function~\eqn{Quadrat}.
Moreover the Fourier transform factors in a similar way. 
If decomposed into partial waves
the Fourier transform reduces to a Fourier-Bessel transform and one obtains
\be \label{factorization}
\Im G(\bk,\bk,E)
=
-4\pi \kappa \sum_l  (2l+1) f_l^-(\kappa,k) \bar f^-_l(\kappa,k)
\ee
with
\be
 f_l^-(\kappa,k)=\int_0^\infty dr \,r^2 f_l^-(\kappa,r)j_l(kr)
\pkt \ee
This Fourier transform is problematic as the integral is not absolutely
convergent.
This problem is analyzed in Appendix C, making use of Wronskian identities
for the free and exact Green functions.
The final result is
\begin{eqnarray}
f_l^-(\kappa,k)&=&\left(e^{2i\delta_\kappa}+1\right)
\frac{\pi}{4k\kappa}\left\{(-1)^{l+1}\delta(k+\kappa)+\delta(k-\kappa)\right\}
\nonumber\\ \label{flmfou}
&&+\frac{\cal{P}}{\kappa^2-k^2}
\int_0^\infty dr r^2 j_l(kr) {\cal {V}}(r)
f^-_l(\kappa,r)\pkt
\end{eqnarray}
Within the range of integration for the parton distribution, Eq.~\eqn{DiaS},
the $\delta$ functions do not contribute as $k^2-\kappa^2 > 0$. So in fact
only the last term contributes, the integral converges as $\calv(r)\to 0$
exponentially as $r \to \infty$. This provides a numerically stable
expression for this Fourier transform.


\section{Numerical computation}
\setcounter{equation}{0}

In the main text we have considered scalar fields in order to
present the scheme for evaluating the parton distributions
on the basis of Green function techniques. We will continue to do so here.
Obviously the real computations were done using the formalism 
based on the Grand spin reduction of the fermion Green function
as presented
in Appendix B, leading to $4\times 4 $ complex coupled differential equations.

\begin{figure}[htbp]
  \begin{center}
    \leavevmode
     \epsfig{file=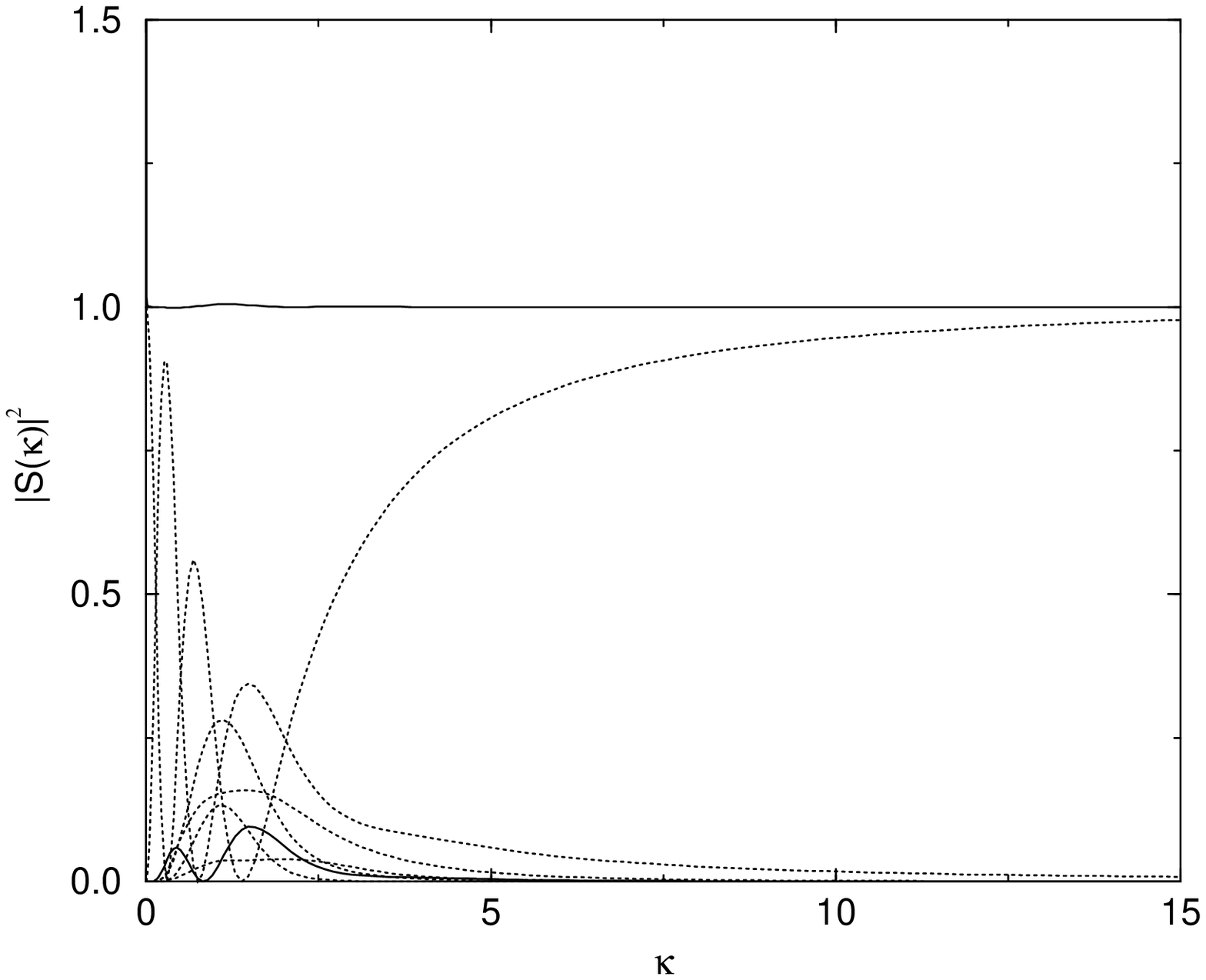,width=13cm,height=7.5cm}  
    {\bf\caption{\label{Figsmatrix}}}
  \end{center}
\end{figure}

The functions $c_l^+(\kappa,r)$ have been calculated by solving the
differential equation
\begin{equation}
\label{fplusDGL}
\left[-\frac{d^2}{d r^2}-2\left(\frac{1}{r}+\kappa 
\frac{h^{(1)'}_{l}(\kappa r)}{h^{(1)}_{l}(\kappa r)}\right)\frac{d}{d r}
\right]c_{l}^
{+}(\kappa ,r)=
-{\cal{V}}(r)\left[1+c_{l}^{+}(\kappa ,r)\right]\pkt
\end{equation}
which follows from Eq.~\eqn{PartDGL} by inserting the ansatz~\eqn{general1}.
We have used a simple four-step Runge-Kutta-scheme, the integration was started
at values of $R$ much larger than the range of the potential, where
$c_l^+$ vanishes by definition, i.e., by its boundary condition.
 The accuracy of these solutions was checked by the
Wronskian relation, which was constant to at least six significant
digits. As a further check we used the unitarity of the $S$ matrix as 
the $4\times 4$ matrix relation~\eqn{unitar}. This is displayed in
Fig.~\ref{Figsmatrix} 

\begin{figure}[htbp]
  \begin{center}
    \leavevmode
     \epsfig{file=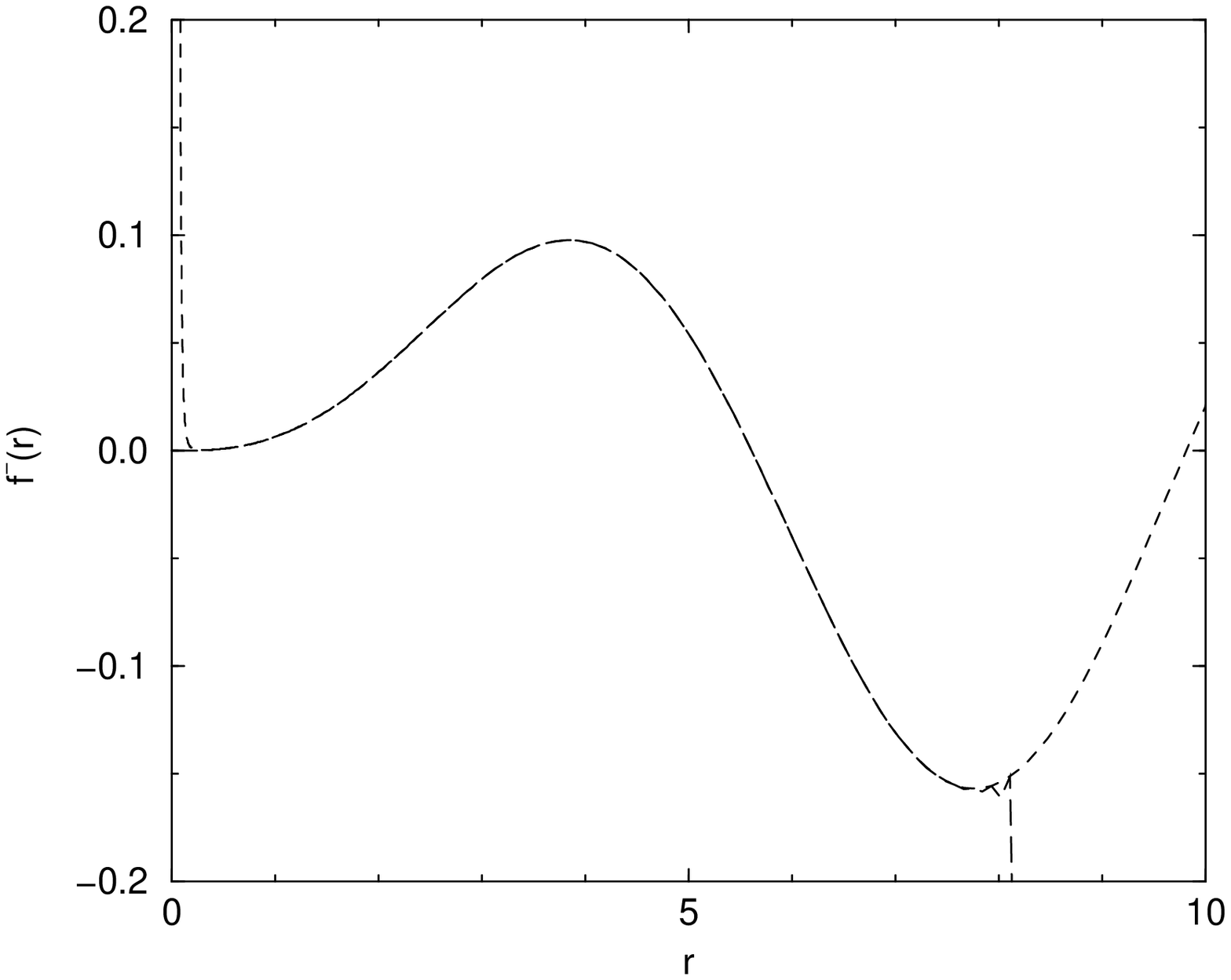,width=13cm,height=7.5cm}  
    {\bf\caption{\label{Figoverlap}}}
  \end{center}
\end{figure}

The functions $f_l^-$ were then obtained in the form
\eqn{first}. As the Hankel functions become singular
at $r=0$ this form is not suitable at very small $r$. In this region
we have used the form~\eqn{general2}, starting the integration
of the differential equation at $r=0$. This solution was 
normalized correctly by fitting it to the composed solution
at some value or $r$ where both solutions are reliable. This is
displayed (for one component of the real $4\times 4$ solution) 
in Fig.~\ref{Figoverlap}.

\begin{figure}[htbp]
  \begin{center}
    \leavevmode
     \epsfig{file=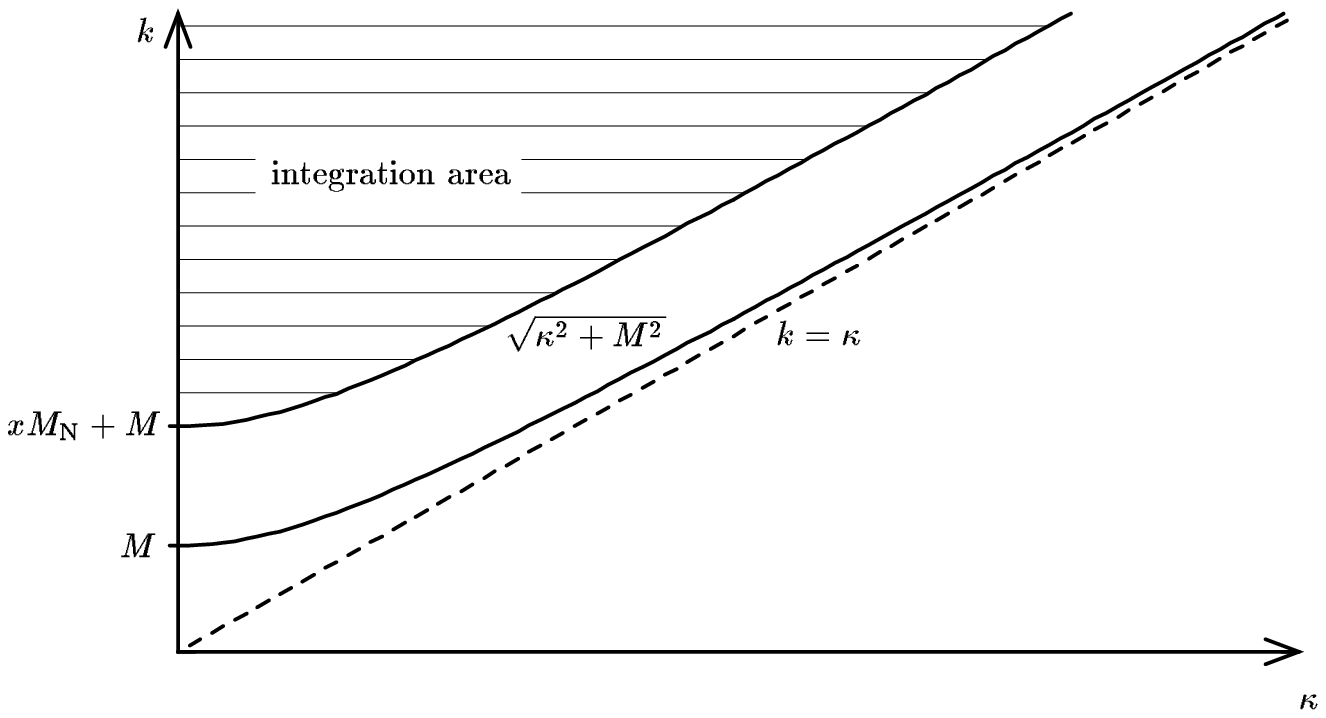,width=13cm,height=7.5cm}  
    {\bf\caption{\label{Figkinematischerplot}}}
  \end{center}
\end{figure}

The next step in the computation of $f_l^-(\kappa,k)$, i.e.,
the Fourier-Bessel transform, using Eq.~\eqn{flmfou}. The Bessel functions
were generated recursively, for small $r$ they were obtained 
via power series expansion. 
The integrand is then obtained by squaring the Fourier transform and by
supplying the  pole prefactor $1/(\kappa^2-k^2)^2$. In the fermion case
the corresponding operation is given in Eq.~\eqn{end}.
In this way the integrand is evaluated on a lattice in the
variables $k$ and $\kappa$. 
The region of integration is displayed in Fig.~\ref{Figkinematischerplot};
the dependence on the scaling variable $x$ comes from the
boundary $k \geq k_{\rm min} =|xM\sub{N}-k_0|$ with 
$k_0=-\sqrt{\kappa^2+M^2²}$ and some
simple prefactors, see Eq.~\eqn{traceS}. The first step is the
integration over $k$. As we work with a discrete lattice
the lower end of integration could lead to a step structure; this
is avoided by fitting the behavior near  $k_{\rm min}$ down to this boundary. 
The integration over large $k$ is unproblematic, both due
to the prefactor $1/(\kappa^2-k^2)^2$ as to the decrease of the
squared Fourier transform.

\begin{figure}[htbp]
  \begin{center}
    \leavevmode
     \epsfig{file=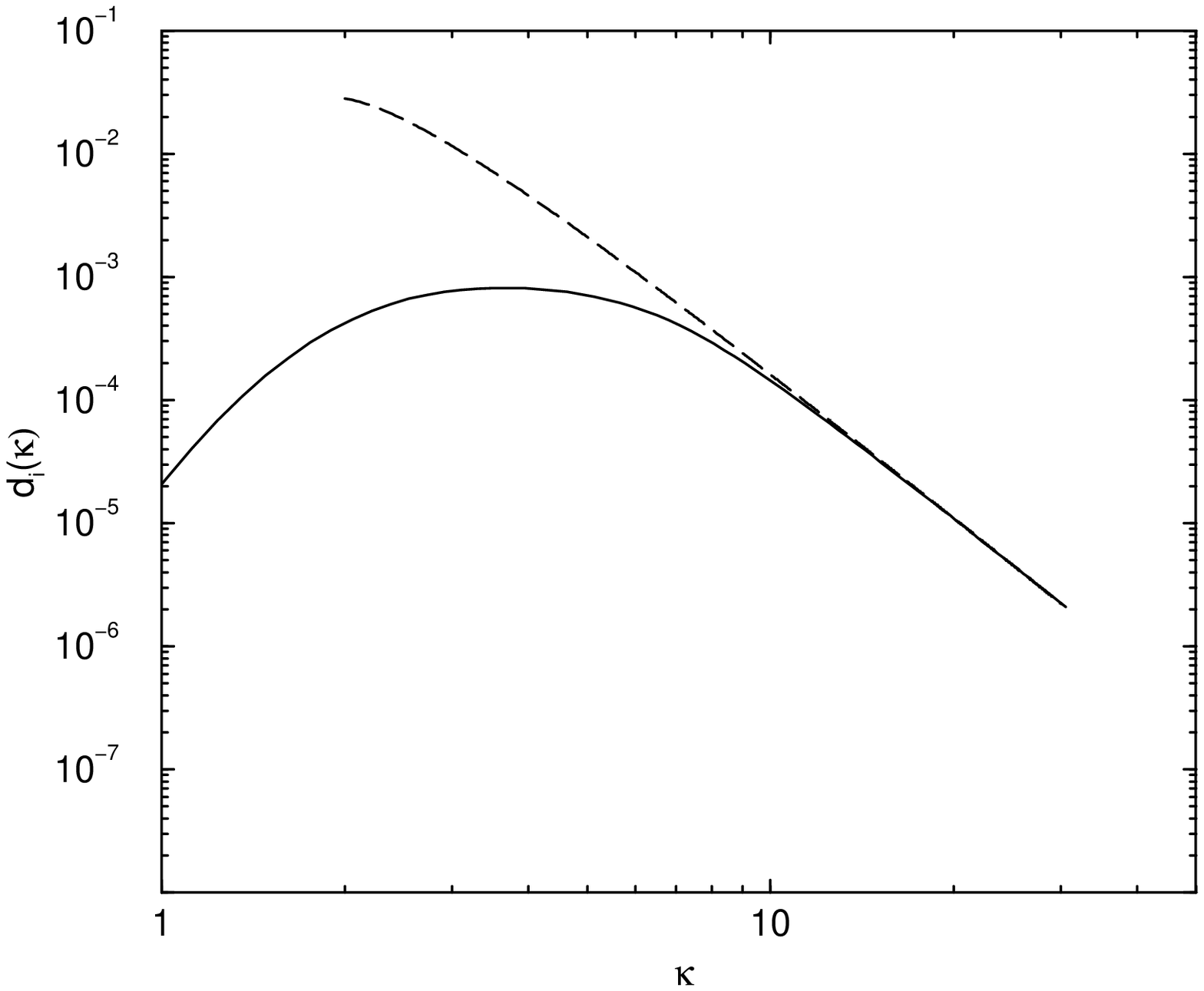,width=13cm,height=7.5cm}  
    {\bf\caption{\label{Figkappaintegrand}}}
  \end{center}
\end{figure}

The next step is the $\kappa$ integration. Here the lower boundary
is  $\kappa=0$ and the integrand is well-behaved there. The 
behavior at large $\kappa$ is powerlike, as expected in a loop
integration. The integration
over the computed integrand was extended to
values $\kappa \simeq (25 \div 35) M$, depending on $K$.
 The integrand decreases as $\kappa^{-4}$, so the
integration up to $\infty$ is appended using an 
asymptotic tail of the form $A \kappa^{-4}+B \kappa^{-5}$.
An example is displayed in Fig.~\ref{Figkappaintegrand}. 
This integration over $\kappa$ has to be performed 
at each value of  $x$.

\begin{figure}[htbp]
  \begin{center}
    \leavevmode
     \epsfig{file=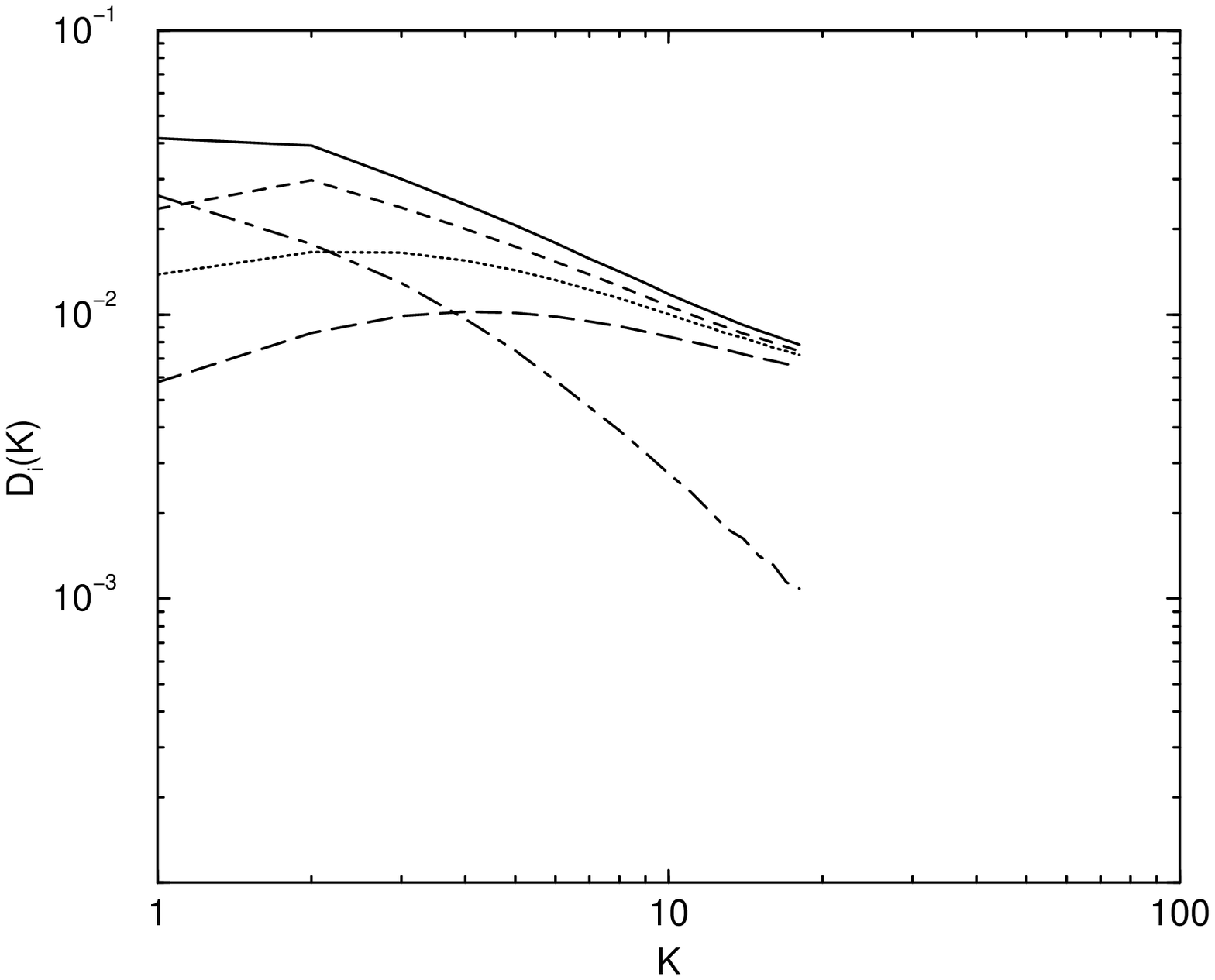,width=13cm,height=7.5cm}  
    {\bf\caption{\label{FigKsumme}}}
  \end{center}
\end{figure}

The final step is the summation over Grand spin $K$ and over parity $P$. 
The behavior of the terms in this sum is displayed in Fig.~\ref{FigKsumme}.
We plot the terms for the different parities for the physical
quark mass $M$ and for the cutoff mass $\Lambda$. 
The  $K^{-1}$ behavior reflects the divergence to be expected from
second order perturbation theory. The difference of the quark mass and
cutoff is taken in the form \cite{Weiss:1997rt}
\be
D^K_i(x)= D^K_i(x,M) - \frac{M^2}{\Lambda^2}  D^K_i(x,\Lambda)
\ee
These terms decrease as $K^{-3}$.
The sum over the $K$-spin was extended to $K=18$. The sum over
the higher angular momenta was appended using a power fit to
the terms computed numerically.

We plot the final results for the isoscalar unpolarized
distribution in Figs.~\ref{Figstructurefunctions1} and
\ref{Figstructurefunctions2}. The antiquark 
distribution is obtained as $\bar D_i(x)=-D_i(-x)$

\begin{figure}[htbp]
  \begin{center}
    \leavevmode
     \epsfig{file=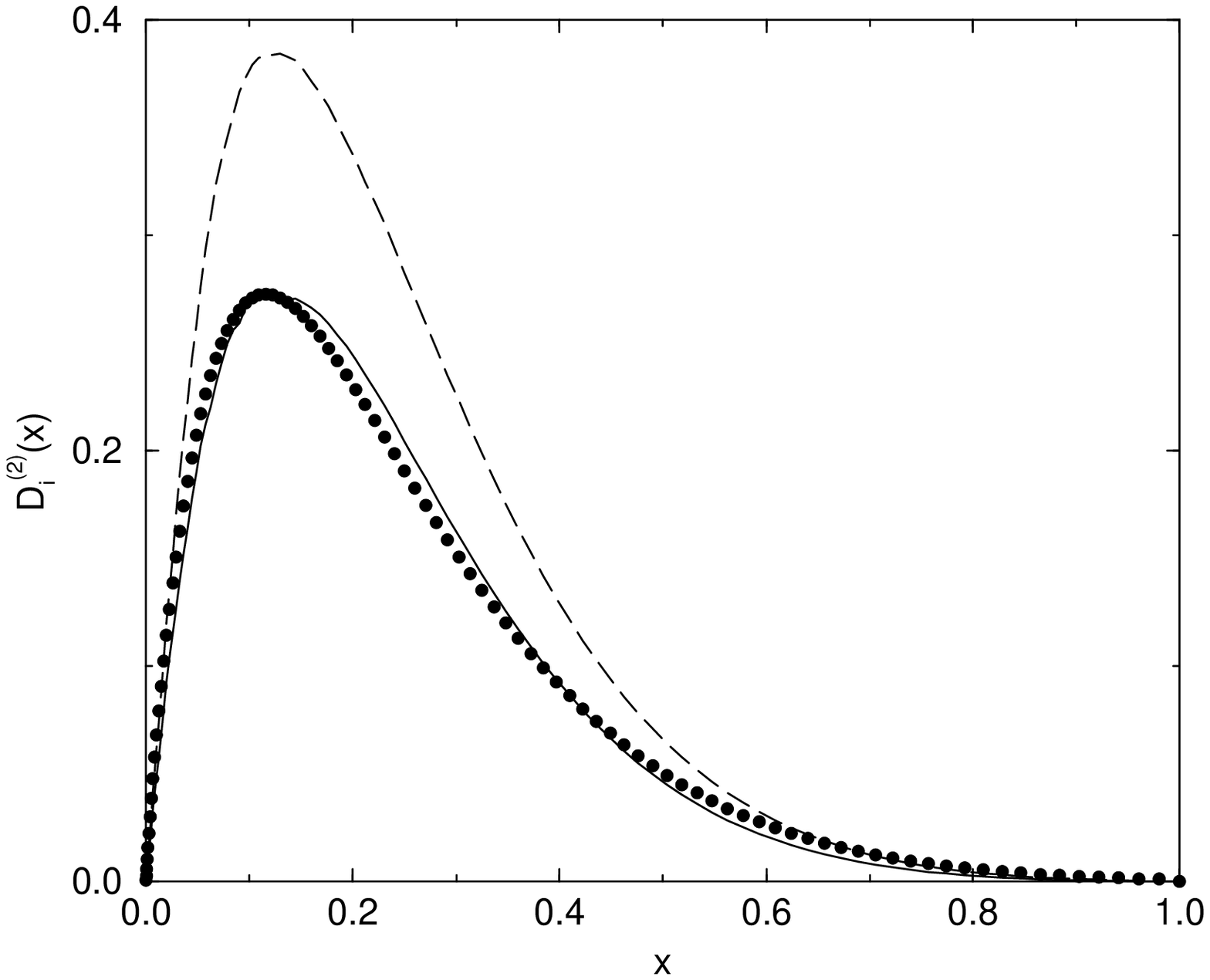,width=13cm,height=7.5cm}  
    {\bf\caption{\label{Figsecorder}}}
  \end{center}
\end{figure}

\begin{figure}[htbp]
  \begin{center}
    \leavevmode
     \epsfig{file=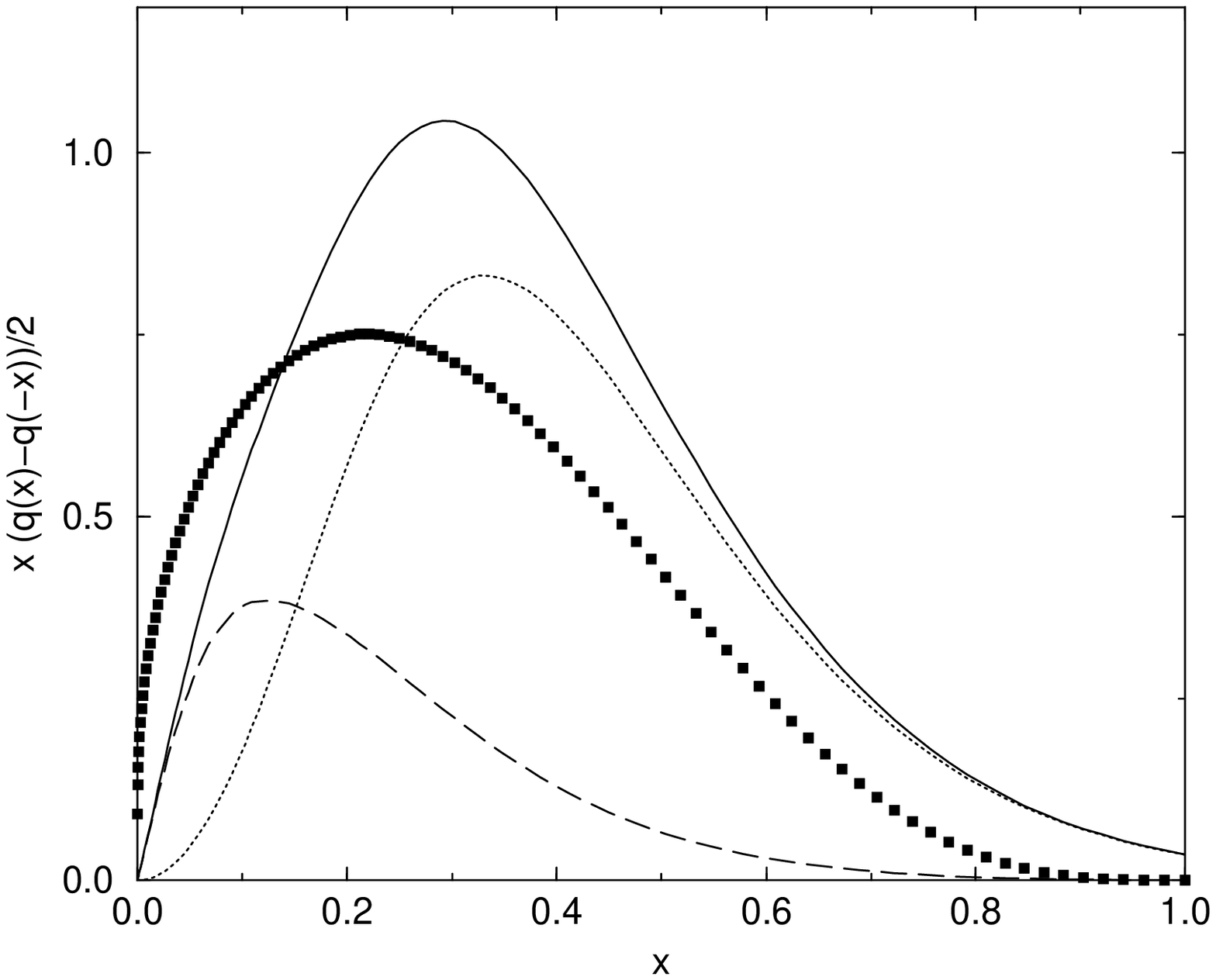,width=13cm,height=7.5cm}  
    {\bf\caption{\label{Figstructurefunctions1}}}
  \end{center}
\end{figure}

\begin{figure}[htbp]
  \begin{center}
    \leavevmode
     \epsfig{file=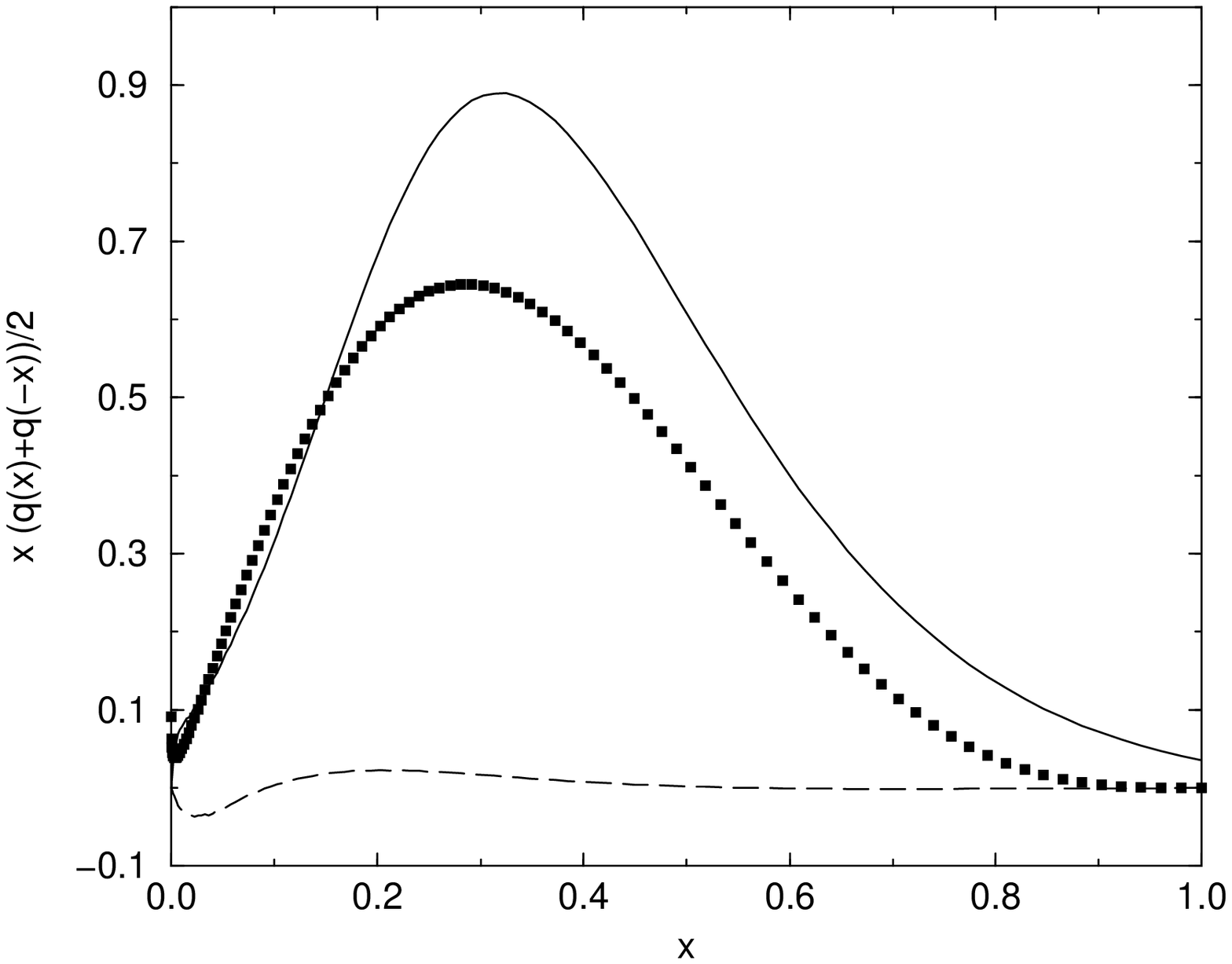,width=13cm,height=7.5cm}  
    {\bf\caption{\label{Figstructurefunctions2}}}
  \end{center}
\end{figure}

It is interesting to compare these results with 
second order perturbative contribution
\begin{eqnarray}
\label{secondorder}
D_{i}(x)&=& \frac{N\sub{c}M\sub{N}M^2}{4\pi^2}
\int\frac{d^3{\bk}}{(2\pi)^3}\theta(k^3-|x|M\sub{N})
\Tr\left(\tilde U({\bk})[\tilde U({\bk})]^\dagger\right)
\nonumber\\
&&
\times \left[\ln\frac{\Lambda^2+\kappa^2}{M^2+\kappa^2}-
\frac{\kappa^2(\Lambda^2-M^2)}{(\lambda^2+\kappa^2)(M^2+\kappa^2)}\right]
\end{eqnarray}
with 
\be
\kappa^2=\frac{|x|M\sub{N}(k^3-|x|M\sub{N}){\bk}^2}{(k^3)^2}\pkt
\ee
This is done for the sum of quark and antiquark distributions
in Fig.  \ref{Figsecorder}. It is a further numerical cross check
to compute the second order perturbative contribution
form the mode expansion. This can be done by replacing in the
expression for the Fourier transform of the mode 
function $f_m^{\alpha-}(\kappa,k)$, Eq. \eqn{fmalphafou},
the exact solution $f_n^{\alpha -}(\kappa,r)$ by the
free solution $\delta_n^\alpha j_{K_n}(\kappa r)$. This comparison
is included as well in Fig.\ref{Figsecorder}. The agreement is not perfect,
but it proves the overall consistency and illustrates the precision
of our results. As the higher order perturbative contributions,
and therefore their error, are less important this graph can be
used to improve the overall result by replacing the computed second order by
the analytic one.  

Finally, it was also checked that the relation
\begin{equation}
\int_0^1 ( D_i(x)+D_i(-x)) dx=N\sub{c}
\end{equation}
was fulfilled within $0.1\%$. In principle this relation 
\cite{Diakonov:1996sr} is valid exactly 
 if the integration is extended to $x=\infty$. As seen
in the figures, the integrand is already very small at $x=1$.


\section{Results and Conclusions}
\setcounter{equation}{0}

We have presented here a new evaluation of the isoscalar nucleon 
structure functions, using the continuum approach based on
the quark Green function in an external background field. 
This constitutes a natural extension of our use of 
Green function methods to the computation
of self-consistent solutions \cite{Baacke:1998nm}.
The method presents various internal cross checks as 
mentioned in the previous section. It is a relatively fast
method, the computation of the structure function
takes about $5$ hours on a standard PC with a $450$ MHz
Pentium processor. 

The self-consistent solution was computed using our previous methods.
For the quark-meson coupling we have chosen
the value $g=4$ which was found to be preferred by the comparison of the
predicted static parameters with experiment. For $f_\pi$ we have taken
the experimental value $93$ MeV.

The results for the parton distributions are shown in 
Figs.~\ref{Figstructurefunctions1} and
\ref{Figstructurefunctions2}. 
As compared to previous analyses, based on the explicit wave functions
of a discretized system, we find an overall agreement as far as order
of magnitude and typical structure are concerned. 
The results are similar to those of
Weiss and Goecke \cite{Weiss:1997rt} and of
Wakamatsu and Kubota \cite{Wakamatsu:1998en}. 
There is a tendency, however,
of our structure functions to peak at somewhat smaller values of $x$.
In view of some differences in the self-consistent hedgehog profiles 
and parameters this is however a very satisfactory agreement 
of entirely different numerical approaches.

Our results are compared  to the parton distributions 
obtained in the  GRV parametrisation\cite{Gluck:1995uf} 
at $Q^2=0.40$ GeV$^2$, that is, of the order  of the cutoff $\Lambda^2$.
The agreement is certainly qualitative and even semi-quantitative.

One of our main results that our method corroborates previous analyses 
in a very satisfactory way. However, it also adds a new and efficent tool
for such and similar computations. As we have shown it comprises
various internal cross checks, some of which are due the
use of analytical methods of potential scattering.
Furthermore the method is  comparatively fast, and so can be useful
for more elaborate computations.

The most obvious continuation of the present work would be the
computation of the isovector structure functions.
As it is well-known these functions vanishe in order $N_C^{(1)}$, and
the first nonvanishing part is of the order $N_C^{(0)}\sim \Omega^1$. 
These $N_C^{(0)}$ parts are necessary in order to obtain
the $u$- and $d$- quark distributions separately.
In analogy to the computation of the momentum of 
inertia\cite{Baacke:1998nm} (first done
by Wakamatsu and Yoshiki \cite{Wakamatsu:1991ud} using discretization),
it is feasible to extend our method to $N_C^{(0)}$-distribution functions,
taking into account the rotation of the soliton.

 The fundamental formulae have been found after a long discussion
in \cite{Weigel:1997b}, \cite{Wakamatsu:1998en} 
and \cite{Pobylitsa:1998tk}.
Using the same procedure as for the $N_C^{(1)}$-case, they can 
again be 
written in terms of Minkowskian Green functions.
One obtains 
\begin{eqnarray}
  \left[u(x)-d(x)\right]^{(1)}&=&-2T_3\frac{N_{\rm C}M_{\rm N}}
{24\pi I}{\rm Im} 
\int \frac{d^4k}{(2\pi)^4}\int \frac{d^3k'}{(2\pi)^3}\delta(k_0+k_3-xM_{\rm N})
\nonumber\\&&\times{\rm Tr} S_{\rm F}(k,k',k_0)\tau_a\gamma^0 S_{\rm F}
(k',k,k_0)
\tau_a(\gamma^0+\gamma^3)\Theta(-k^0)
\end{eqnarray}
and 
\begin{eqnarray}
  \left[u(x)-d(x)\right]^{(2)}&=&-2T_3\frac{N_{\rm C}}{4 I M_{\rm N}}
\frac{\partial}{\partial x} \left[u(x)+d(x)\right]\,,
\end{eqnarray}
The computation of the fermion Green functions in the external hedgehog field 
has been outlined here. Comparing to the computation of the 
isosinglet structure functions we need one more integration, the one
over $\bk'$. In order to avoid Fourier transforming it can be done in
$x$ space. Furthermore one looses the factorization property
\eqn{factorization}. On the other hand the convergence in momentum
and angular momentum will be faster, as for the analogous leading
order Feynman graph. Furthermore the functions themselves are
smaller, so one can somewhat reduce the numerical accuracy.
So this computation seems feasible and is being undertaken at present.     

A further issue will be the computation  of 
the polarized structure 
functions \cite{Wakamatsu:1998en,Diakonov:1997vc} and of 
the skewed parton distributions
\cite{Petrov:1998kf}. It would be interesting, furthermore, to study the full
$Q^2$ dependence, as has been done for the bound state
contribution by Weigel, Gamberg and Reinhard 
\cite{Weigel:1997jh} as well as by Ruiz Arriola \cite{RuizArriola:1998er}.

As a side result we have obtained a very fast method 
for calculating the phase shift. Alternatively to the method
employed in \cite{Moussallam:1989uk,Farhi:2000ws}, the computation of the
system of smooth nonoscillating functions $c_n^{\alpha+}$ 
 allows the extraction of the
multichannel $S$ matrix.
A further possibility - and cross check - is the use of \eqn{phasesumrule}
and its multichannel extension.

\vspace{2cm}

\noindent{\Large\bf{Acknowledgments}}
\vspace{0.5cm}

\noindent 
H.S. acknowledges the support
by the Graduiertenkolleg ``Er\-zeug\-ung und Zerf\"alle von 
Elementar\-teil\-chen''.

\begin{appendix}
\section{Parton distribution}
\setcounter{equation}{0}
The fermion Green function has been formally expressed by the eigenmodes
$U_\alpha$  and $V_\alpha$ in Eq.~\eqn{greenuv}. Its Fourier transform
\eqn{greenfou} is given by
\bea \label{greenuvfou}
S_F(\bk,\bk',k_0)=
-\sum_{\alpha >0}\frac{U_\alpha(\bk)\bar U_\alpha(\bk')}
{k_0-E_\alpha+i\epsilon}
-\sum_{\alpha < 0}\frac{V_\alpha(\bk)\bar V_\alpha(\bk')}
{k_0+|E_\alpha|-i\epsilon}
\eea
where the Fourier transform for the eigenspinors is defined via
\bea
V_\alpha(\bx)&=&\int\frac{d^3k}{(2\pi)^3}V_\alpha(\bk)e^{i\bk\cdot\bx}
\\
U_\alpha(\bx)&=&\int\frac{d^3k}{(2\pi)^3}U_\alpha(\bk)e^{i\bk\cdot\bx}
\pkt\eea
The parton distribution of sea quarks
is given by \cite{Diakonov:1996sr}
\be
q(x)=\frac{N\sub{c} M\sub{N}}{2\pi}
\int d^3X\int_{-\infty} ^\infty dz^0e^{ixM\sub{N}z^0 }
\sum_{\alpha < 0}e^{i|E_\alpha|z^0}
V^\dagger_\alpha(-\bX)(1+\gamma_0\gamma^3)V_\alpha(-\bX-z^0\bn_3)
\pkt\ee
If one rewrites this in terms of the Fourier transforms,
the $z^0$ and $\bX$ integration can be performed with the result
\be
q(x)=N\sub{c} M\sub{N}\int\frac{d^3k}{(2\pi)^3}
\sum_{\alpha < 0}\delta(|E_\alpha|+xM\sub{N}-k^3)
\tr\left[(1+\gamma_0\gamma^3)V_\alpha(\bk)V^\dagger_\alpha(\bk)\right]\pkt
\ee
We note that $1+\gamma_0\gamma^3$ is a hermitean operator.
Considering the Green function in the form~\eqn{greenuvfou}
we see that
\bea
&&S_{\rm F}(\bk,\bk,k_0)\gamma^0-\gamma^0S^\dagger_F(\bk,\bk,k_0)\nonumber
\\
&&= \sum_{\alpha >0}2 \pi \delta(k^0-E_\alpha)U_\alpha(\bk)U^\dagger_\alpha(\bk)
-\sum_{\alpha< 0}2 \pi \delta(k^0+|E_\alpha|)V_\alpha(\bk)V^\dagger_\alpha(\bk)
\pkt
\eea
We therefore find
\be
q(x)=-\Im N\sub{c} M\sub{N}\int\frac{d^4k}{(2\pi)^4}
\delta(xM\sub{N}-k^3-k^0)\Theta(-k^0)
\tr\left[(\gamma_0+\gamma^3)S_{\rm F}(\bk,\bk,k^0)\right]\pkt   
\ee
The restriction of the integration to negative $k^0$ does not appear
in (A.9) of \cite{Diakonov:1996sr}. It does appear in their
Eq. (7.3) for the second order perturbative contribution.
 As is apparent here it is crucial for 
restricting the integration to  the sea quark states.
We have remarked below Eq.~\eqn{flmfou} that $\kappa^2 -k^2 > 0$
so that the $\delta$ functions do not contribute.
In terms of the variables used here 
\be
k^2- \kappa^2= \bk^2-(k^0)^2+ M^2 \geq (k^3)^2+ M^2 - (k^0)^2
=  M^2 + x^2 M\sub{N}^2 - 2 x M\sub{N} k^0 > 0
\pkt
\ee

The antiquark distribution is determined formally by the positive energy
eigenspinors. This means that one has to replace the $\Theta$ function
by $\Theta(k_0)$. At the same time the energy of these levels has to
be replaced by $-E_\alpha,\, \alpha > 0$. In the Green function
this reversal of sign is obtained by replacing $x$ by $-x$.


\section{The  Green function for the quark system}
\setcounter{equation}{0}
\subsection{The  boson Green function}
As we have described in section 2 the fermion Green function
has been reduced to a bosonic one which satisfies the
differential equation~\eqn{DGL2}. It is a $8\times 8$
matrix. Using parity it becomes block diagonal with
two $4 \times 4$ matrices. 
As the system is invariant with respect to $K$ spin, 
it can be expanded \cite{Kahana:1984} 
with respect to K-spin harmonics $\Xi^{K,K_z}_n$ via
\be
G(\bz,\bz',E)=\sum_{K,K_z,P}
g^{K,P}_{mn}(r,r',\kappa) \Xi^{K,K_z}_m(\hbz)
\otimes\Xi^{K,K_z \dagger}_n(\hbz')
\ee
with $\kappa=\sqrt{E^2 -M^2}$.
The radial Green functions are $ 4\times 4$ matrices.
They can be written in terms of $4$ - component mode functions
as
\bea\label{gtheta}
g_{mn}(r,r',\kappa)&=&\kappa\left[
            \theta(r-r')f^{\alpha +}_{m} (\kappa,r)f^{\alpha -}_{n}(\kappa,r') 
\right.
\\ \nonumber&&\left.
           +\theta(r'-r)f^{\alpha -}_{m} (\kappa,r)f^{\alpha +}_{n}(\kappa,r') 
            \right]\pkt
\eea
The mode functions satisfy the differential equation
\be
\left[-\frac{d^2}{d r^2}-\frac{2}{r}\frac{d}{d r}
+\frac{K_n(K_n+1)}{r^2}-\kappa^2\right]f^\alpha_n(\kappa,r) =
-{\cal{V}}_n^{n'}(r)f_{n'}^\alpha(\kappa ,r)\pkt
\ee
The subscript denotes the four components, the superscript
$\alpha=1\dots 4$ labels the four independent solutions 
which form a fundamental system.
The potential $\calv_n^{n'}(r)$ is the grand spin partial wave reduction
of ${\bf \calv (\bx)}$, Eq.~\eqn{potential}. 
Its explicit form has been given, e.g.,
in Appendix A of \cite{Baacke:1998nm}.

In order to split off the behavior at $r=0$ and at $r=\infty$
of the free solutions we write them in the form  
\bea
f_{m}^{\alpha\,+}(\kappa ,r)&=&\left[\delta_m^\alpha
+c_{m}^{\alpha\,+}(\kappa ,r)\right]h^{(1)}_{K_m}(\kappa r)\kma
\\
f_{m}^{\alpha\,-}(\kappa ,r)&=&\left[\delta_m^\alpha
+c_{m}^{\alpha\,-}(\kappa ,r)\right]j_{K_m}(\kappa r)\pkt
\eea
Here $K_m$ takes the values $K-1,K,K,K+1$ for the
four components labelled by $m$.
As for the single component case presented in section 3
the boundary conditions are
\bea
\lim_{r\to \infty}c_m^{\alpha+}(\kappa,r) &=& 0\kma
\\
\lim_{r\to 0}c_m^{\alpha-}(\kappa,r) &=& {\rm const.}
\kma
\eea
and the fundamental system $f_m^{\alpha-}$ is fully specified
by the Wronskian
\begin{equation}
r^2 W^{\alpha\beta}(f^+,f^-)=
r^2\left(f_m^{\alpha\, +}\frac{d}{d r}f_m^{\beta\, -}-
f_m^{\beta\, -}\frac{d}{d r}f_m^{\alpha\, +}\right)=
\frac{-i}{\kappa}
\delta^{\alpha\beta}\pkt
\end{equation}
As for the single-component case $f_n^{\alpha-}$ can be written as
\begin{eqnarray}
\label{lc4}
f_{m}^{\lambda\,-}(\kappa ,r)&=&\frac{1}{2}
\Biggl\{
\left(\delta^\lambda_\alpha+\bar{c}^{\lambda +}_\alpha(\kappa,0)\right)
\left[\left(\delta+{c}^{+}(\kappa,0)\right)^{-1}\right]^\alpha_\gamma
\left(\delta^\gamma_m+{c}_{m}^{\gamma\, +}(\kappa ,r)\right)
h^{(1)}_{K_m}(\kappa r)\nonumber\\&&+
\left(\delta^\lambda_m+\bar{c}_{m}^{\lambda\, +}(\kappa ,r)\right)
h^{(2)}_{K_m}(\kappa r)\Biggr\}
\end{eqnarray}
and it was computed in this form except at very small $r$.
The $S$ matrix is defined by the asymptotic behavior of
the regular solution
 $f_m^{\lambda\, -}(\kappa ,r)$, from the representation~\eqn{lc4}
we find 
\begin{eqnarray}
\label{Sm}
S_m^\lambda=
\left({1+\bar c^+(\kappa ,0)}\right)_\alpha^\lambda
\left(\left[{1+ c^+(\kappa ,0)}\right]^{-1}\right)^\alpha_m\pkt
\end{eqnarray}
From this form it is not obvious that it is unitary.
We have to use the further information that the Green function
forms a symmetric matrix
\begin{eqnarray}\label{gsymm}
 g_{mn}(R,R,\kappa)&=&g_{nm}(R,R,\kappa)
\end{eqnarray}
which leads to  the relation
\begin{eqnarray}
\label{Sm4}
\left(1+\overline{c}(\kappa,0)\right)^n_\alpha
\left[\left(1+{c}(\kappa,0)\right)^{-1}\right]^\alpha_m
=\left(1+\overline{c}(\kappa,0)\right)^m_\alpha
\left[\left(1+{c}(\kappa,0)\right)^{-1}\right]^\alpha_n
\pkt\end{eqnarray}
Using this relation one obtains
\begin{eqnarray} \label{unitar}
S_m^\lambda {S^*}_n^\lambda
&=&\left(1+\bar{c}^+(\kappa ,0)\right)^\lambda_\alpha
\left[\left(1+{c}^+(\kappa ,0)\right)^{-1}\right]^\alpha_m
\left(1+{c}^+(\kappa ,0)\right)^\lambda_\beta
\left[\left(1+\bar{c}^+(\kappa ,0)\right)^{-1}\right]^\beta_n\nonumber\\
&=&\left(1+\bar{c}^+(\kappa ,0)\right)^\lambda_\alpha
\left[\left(1+\bar{c}^+(\kappa ,0)\right)^{-1}\right]^\alpha_m
\left(1+{c}^+(\kappa ,0)\right)^\lambda_\beta
\left[\left(1+{c}^+(\kappa ,0)\right)^{-1}\right]^\beta_n\nonumber\\
&=&\delta_{mn}\pkt
\end{eqnarray}
This relation can be used to check the numerics. The inaccuracy is less than
$10^{-6}$.
Using the symmetry relation~\eqn{gsymm} in the form~\eqn{Sm4}
one can show that the imaginary part of the Green function 
again factorizes:
\be
\label{end}
\Im g_{mn}(r,r',\kappa)=-\kappa f_m^{\alpha-}(\kappa,r)
f_n^{\alpha-}(\kappa,r')
\pkt\ee

For the numerical computation one needs the Fourier transform
of the mode functions $f_m^{\alpha -}$. It is given by
\begin{eqnarray} \label{fmalphafou}
f_m^{\alpha-}(\kappa,k)&=&
\left(e^{2i\delta(\kappa)}+1\right)_m^\alpha
\frac{\pi}{4k\kappa}\delta(k-\kappa)
\left\{(-1)^{m+1}\delta(k+\kappa)+\delta(k-\kappa)\right\}
\nonumber\\&&+\frac{\cal{P}}{\kappa^2-k^2}
\int_0^\infty dr r^2 j_{K_m}(kr) {\cal{V}}_{mn}(r)
f^{\alpha-}_n(\kappa,r)
\end{eqnarray}
where $e^{2i\delta(\kappa)}+1$ means
\begin{eqnarray}
\left(e^{2i\delta(\kappa)}+1\right)_\lambda^\alpha=S_\lambda^\alpha+
\delta_\lambda^\alpha\pkt
\end{eqnarray}

With these preparations we can obtain the Green function for the fermion system. 
We use  the Fourier transform of Eq.~\eqn{DGL2}: 
\begin{eqnarray}
\label{SiG}
S(\bk,\bk,\kappa)&=&
\left(-\sqrt{\kappa^2+M^2}-k \balpha \hbk + \gamma_0 M\right)
G(\bk,\bk,\kappa)\,.\nonumber\\
\end{eqnarray}
The action of $\balpha {\hbk}$ onto the spinor harmonics is given by
\begin{eqnarray}
\label{omega1}
\bsigma\hbk
\left(\begin{array}{l}\Xi_{1}(k)\\ \Xi_{2}(k)\\ \Xi_{3}(k)\\
\Xi_{4}(k)\end{array} \right)&=&
-\left(\begin{array}{l}\Xi_{2}(k)\\ \Xi_{1}(k)\\ \Xi_{4}(k)\\
\Xi_{3}(k)\end{array} \right)\pkt
\end{eqnarray}
Combining the $\Xi$ spinors, the $\gamma$ matrices and Eq.~(\ref{SiG}) 
one finds after some algebra
\begin{eqnarray}
\nonumber
\int d\Omega_{\bk}\,  \Im\tr \left[\left(\gamma_0+\gamma_3\right)
S(\bk,\bk,\kappa)\right]&=&\Im
\sum_{K,P}(2K+1)\Biggl\{
\biggl(2E-xM\sub{N}+M\biggr)\biggl(g_{11}+g_{44}\biggr)\\
&& \label{traceS}
-\biggl(2E-xM\sub{N}-M\biggr)\biggl(g_{22}+g_{33}\biggr)\\&&
-2(xM\sub{N}-E)\frac{E}{k}\biggl(g_{12}+g_{34}\biggr)
+2k\biggl(g_{12}+g_{34}\biggr)\nonumber\pkt
\Biggr\}\,
\end{eqnarray}
Here we have suppressed  labels and arguments  of $g_{ij}$; explicitly
they  are given by
\begin{eqnarray}
\Im  g^{K,P}_{ij}(k,k,\kappa)&=&-
\frac{\kappa}{\left(\kappa^2-k^2\right)^2}
\int_0^\infty dr r^2 j_{K_i}(kr) {\cal{V}}_{in}(r)
f^{\alpha-}_n(\kappa,r)
\nonumber\\
&&\times\int_0^\infty dr r^2 j_{K_j}(kr) {\cal{V}}_{jn'}(r)
f^{*\alpha-}_{n'}(\kappa,r)\,.
\end{eqnarray}


\section{Fourier transform}
\setcounter{equation}{0}
We here discuss the Fourier transform
of the mode function $f^-_l(\kappa,r)$
\be
 f_l^-(\kappa,k)=\int_0^\infty dr \,r^2 f_l^-(\kappa,r)j_l(kr)
\pkt \ee
This Fourier transform is problematic numerically as the integral
is not absolutely convergent.

We start with the differential
equations for $f_l^-(\kappa,r)$ given in Eq.~(\ref{PartDGL}) and for $j_l(kr)$
which reads
\begin{eqnarray}
\label{DGLj}
-\frac{1}{r^2}\frac{d}{dr}r^2\frac{d}{dr}j_l(k r)
+\frac{l(l+1)}{r^2}j_l(k r)&=&
k^2j_l(k r)
\pkt\end{eqnarray}
From these we derive
\begin{eqnarray}
&&\nonumber
-\int_0^R dr \left(j_l(k r)\frac{d}{dr}r^2\frac{d}{dr}f_l^-(\kappa,r)-
f_l^-(\kappa,r)\frac{d}{dr}r^2\frac{d}{dr}j_l(k r)\right)\\
&&
+\int_0^R dr r^2 j_l(k r) {\cal {V}}(r) f_l^-(\kappa,r)=
\left(\kappa^2-k^2\right)
\int_0^R dr r^2 j_l(k r) f_l^-(\kappa,r)
\end{eqnarray}
and after  integration by parts 
\begin{eqnarray}
\label{ill}
\int_0^R dr r^2 j_l(kr)f^-_l(\kappa,r)&=&\frac{1}{\kappa^2-k^2}\Biggl\{
\int_0^R dr r^2 j_l(kr) {\cal {V}}(r)f^-_l(\kappa,r)\\&&
-R^2\left[j_l(kR)\frac{d}{dR}f^-_l(\kappa,R)
-f^-_l(\kappa,R)\frac{d}{dR}j_l(kR)\right]\Biggr\}\nonumber\pkt
\end{eqnarray}
The boundary contribution at $r=0$ vanishes, because both $f^-_l$
and $j_l$ behave as $r^l$. The first term on the
right hand side is well-defined in the limit $R\to\infty$, due to the decrease
of the potential. The two  terms in the bracket display an oscillatory behavior
with constant amplitude due to the asymptotic behavior of $f_l^-$ :
\be
f^-_l(\kappa,r)\simeq f^{-,\infty}_l(\kappa,r)=\frac{1}{2}
\left(h^{(2)}_l(\kappa,r)+e^{2i\delta_\kappa}h^{(1)}_l(\kappa,r)\right)\pkt
\ee
This behavior  makes the
limit $R\to \infty$ of the integral on the left hand side ill-defined.
It is these asymptotic oscillations that we have to split off.

Using the Wronski determinants for the free Bessel functions
and Eq.~\eqn{ill} one finds
\begin{eqnarray}
&&\int_0^R dr r^2 j_l(kr)
\left(f^-_l(\kappa,r)-f^{-,\infty}_l(\kappa,r)\right)\nonumber\\
&&=\frac{1}{\kappa^2-k^2}\Biggl\{
\int_0^R dr r^2 j_l(kr) {\cal {V}}(r)
f^-_l(\kappa,r)
\nonumber\\&&
-R^2\biggl[j_l(kR)\frac{d}{dR}
\left(f^-_l(\kappa,R)-f^{-,\infty}_l(\kappa,R)\right)
\nonumber\\&&-\left(f^-_l(\kappa,R)-f^{-,\infty}_l(\kappa,R)\right)
\frac{d}{dR}j_l(kR)\biggr]
-i\frac{k^l}{2\kappa^{l+1}}\left(e^{2i\delta_\kappa}-1\right)
\Biggr\}\pkt
\end{eqnarray}
The limit $R\to \infty$ of this equation is well-defined. One obtains
\begin{eqnarray}
\label{nosing}
\int_0^\infty dr r^2 j_l(kr)
\left(f^-_l(\kappa,r)-f^{-,\infty}_l(\kappa, r)\right)&=&\frac{1}{\kappa^2-k^2}
\biggl[\int_0^\infty dr r^2 j_l(kr) {\cal {V}}(r)
f^-_l(\kappa,r)
\nonumber\\&&
-i\frac{k^l}{2\kappa^{l+1}}\left(e^{2i\delta_\kappa}-1\right)\biggr]\pkt
\end{eqnarray}
Both parts of the integral on the left hand side are still
ill-defined.

In scattering theory this problem is solved by working with
normalizable wave functions obtained by giving small
imaginary parts to the momenta (here $\kappa$) and taking the 
appropriate limit to the real axis. This is of course the origin of the
physical cuts. This is usually done using Jost functions
\cite{deAlfaro:1965}, like
$f_l^+(\kappa,r)$, which asymptotically behave as $\exp(\pm i \kappa r)$.
 However, 
for brevity of presentation, we have preferred to work here with $f_l^-$ directly.
The following results are a heuristic presentation of what we have obtained
by using the meticulous $i\epsilon$ prescriptions.  

The Fourier transforms of the free 
solutions $h^{(1)}_l(\kappa r)=j_l(\kappa r)+iy_l(\kappa r)$ 
are{\footnote{$\cal{P}$ means principle
value}} 
\begin{eqnarray}
  \int_0^\infty dr r^2 j_l(kr)j_l(\kappa r)=\frac{\pi}{2k\kappa}
\left\{(-1)^{l+1}\delta(k+\kappa)+\delta(k-\kappa)\right\}\pkt
\end{eqnarray}
and
\begin{eqnarray}
  \int_0^\infty dr r^2 j_l(kr)y_l(\kappa r)=
k^l\frac{\cal{P}}{(\kappa^2-k^2)\kappa^{l+1}}
\pkt
\end{eqnarray}
One obtains 
\begin{eqnarray*}
f_l^-(\kappa,k)&=&
\left(e^{2i\delta_\kappa}+1\right)
\frac{\pi}{4k\kappa}\left\{(-1)^{l+1}\delta(k+\kappa)+\delta(k-\kappa)\right\}
\nonumber
\\
&&+\frac{\cal{P}}{\kappa^2-k^2}
\int_0^\infty dr r^2 j_l(kr) {\cal {V}}(r)
f^-_l(\kappa,r)\pkt
\end{eqnarray*}
As we have mentioned in the main text, the distribution character of this
result does not play a role in the numerical computation.
The finite integral which appears on the right hand side plays a central
role in evaluating the parton distributions.
Using further the methods of potential scattering we can derive a useful
identity relating this integral to the $S$ matrix.

The integral equation accompanying the differential equation
(\ref{PartDGL}) for the solutions $f^+$ reads
\begin{eqnarray}
\label{xxy}
f_l^+(\kappa,r)&=&-i\kappa \int_0^r d r'r^{'2}j_l(\kappa r')\he(\kappa r)
{\cal{V}}(r')f^+_l(\kappa,r')\nonumber\\&&
-i\kappa \int_r^\infty d r'r^{'2}j_l(\kappa r)\he(\kappa r'){\cal{V}}(r')
f^+_l(\kappa,r')\pkt
\end{eqnarray}
As asymptotically $f_l^+(\kappa,r)$ behaves as $\he(\kappa r)$
one finds
\begin{eqnarray}
\frac{i}{\kappa}&=& \int_0^\infty d r'r^{'2}j_l(\kappa r')
{\cal{V}}(r')f^+_l(\kappa,r')\pkt
\end{eqnarray}
Combining $f^-$ using $f^+$ in the usual form
\begin{eqnarray}
f^-_l(\kappa,r)&=&\frac{1}{2}
\left({f^*}^+_l(\kappa,r)+e^{2i\delta_\kappa}f^+_l(\kappa,r)\right)
\end{eqnarray}
the integral takes for $k=\kappa$  the form
\begin{eqnarray} \label{phasesumrule}
\int_0^\infty dr r^2 j_l(\kappa r) {\cal {V}}(r)
f^-_l(\kappa,r)=\frac{i}{2\kappa}\left(e^{2i\delta_\kappa}-1\right)\pkt
\end{eqnarray}
This relation has been used as a  useful crosscheck.
A further cross check is obtained form the Born approximation to this
relation:
\be
-\kappa \int_0^\infty dr\,r^2 j_l^2(\kappa r)\calv(r)=\delta_l^{(1)}(\kappa)
\ee
which determines the behavior of the phase shifts at large $l$ and at large $\kappa$.

\end{appendix}


\newpage

\section*{Figure captions}
\noindent{\bf Figure 1}: 
Unitarity of $S$ matrix:
$(S^\dagger  S)^1_ 1(\kappa)$ for $K^P=1^+$:
solid line: total matrix element,  other lines:
the eight components $S_1^{*a}S_1^a$, $a= 1\dots 4$,
product of real parts and product of imaginary parts.
\vspace{.5cm}

\noindent{\bf Figure 2}: 
Behavior of $f_1^{1-}(\kappa,r)$ for ($K^P=5^+$, $\kappa=1$),
at small $r$. dashed line: computed via Eq.~\eqn{first},
long dashed line: computed via Eq.~\eqn{general2}.
\vspace{.5cm}

\noindent{\bf Figure 3}: 
Integration area in a $k-\kappa$ plane: the dashed line represents
position of the pole, the lower solid line is the minimal lower limit of
the integration, the upper solid line shows the lower limit for a
positive value of $x$.
\vspace{.5cm}

\noindent{\bf Figure 4}:
Convergence of the $\kappa$ integration: integrand for the mode 
function at $x=.25$ for $K^P=5^+$; solid line: the integrand,
dashed line: power fit with $A\kappa^{-4}+B\kappa^{-5}$.
\vspace{.5cm}

\noindent{\bf Figure 5}: 
Behavior of Grand spin contributions to $D_i(x=.25)$ at large values of $K$: 
solid and dashed lines:  parities $+$ and $-$ with dynamical mass; 
dotted and long-dashed lines:  parities $+$ and $-$ with cutoff mass;
dash-dotted lines: regulated contribution. 
\vspace{.5cm}

\noindent{\bf Figure 6}: 
Second order contribution;
solid line: computed via a Feynman graphs; dashed line:
computed via  second order mode sum.
\vspace{.5cm}

\noindent{\bf Figure 7}: 
The isosinglet unpolarized distribution of quarks and antiquarks
$\frac{1}{2}\left[q(x)+\bar{q}(x)\right]$ 
solid line: total result, 
dotted line: valence contribution, 
dashed line: sea contribution, 
squares: NLO GRV\cite{Gluck:1995uf}
parametrisation
\vspace{.5cm}

\noindent{\bf Figure 8}: 
The isosinglet unpolarized distribution of quarks and antiquarks
$\frac{1}{2}\left[q(x)-\bar{q}(x)\right]$ 
solid line: total result, 
dashed line: sea contribution, 
squares: NLO GRV\cite{Gluck:1995uf}
parametrisation
\vspace{.5cm}



\end{document}